\documentclass[twocolumn,twoside,slac_two]{revtex4}
\usepackage{graphicx}
\usepackage{fancyhdr}
\usepackage{hyperref}
\hypersetup{
    colorlinks=true,
    urlcolor=magenta,
}
\usepackage{dcolumn}
\usepackage{bm}
\usepackage{epstopdf}
\usepackage{subfig}
\usepackage{epsfig}
\usepackage{color}
\usepackage{wasysym}
\usepackage{natbib}
\usepackage{twoopt}
\usepackage{dashrule}
\definecolor{ColorTitle}{cmyk}{0,.88,.77,.40}

\newcommand{\AMS}{AMS-02}

\newcommand{\R}{\ensuremath{\mathcal{R}}}

\newcommand{\p}{\ensuremath{p}}

\newcommand{\He}{He}

\newcommand{\B}{B}
\newcommand{\C}{C}

\newcommand{\Oxy}{O}

\newcommand{\Fe}{Fe}

\newcommand{\BC}{{B}/{C}}

\newcommand{\eplus}{\ensuremath{e^{+}}}

\newcommand{\epfrac}{e\ensuremath{^{+}}/(e\ensuremath{^{-}}\,+\,e\ensuremath{^{+}})}

\newcommand{\pbarp}{{\ensuremath{\bar{p}/p}}}

\newcommand{\BeBe}{\ensuremath{{}^{10}\textsf{Be}{/}{}^{9}\textsf{Be}}}

\usepackage{verbatim} % NT used for comments
\begin{document}
\title{Bayesian analysis of cosmic-ray propagation parameters: \\secondary antiparticles from spatial-dependent diffusion models}
\author{Jie Feng\,$^{1,2}$}
\email{jie.feng@cern.ch}
\author{Nicola Tomassetti\,$^{3,4}$}
\email{nicola.tomassetti@cern.ch}
\author{Alberto Oliva\,$^{5}$}
\email{alberto.oliva@cern.ch}
\affiliation{$^{1}$Institute of Physics, Academia Sinica, Nankang, Taipei 11529, Taiwan;}
\affiliation{$^{2}$School of Physics, Sun Yat-Sen University, Guangzhou 510275, China;}
%\affiliation{$^{3}$Department of Physics and Earth Science, Universit{\`a} di Perugia and INFN, I-06100 Perugia, Italy}
\affiliation{$^{3}$Universit{\`a} degli Studi di Perugia \& INFN-Perugia, I-06100 Perugia, Italy;}
\affiliation{$^{4}$LPSC, Universit\'e Grenoble-Alpes, CNRS/IN2P3, F-38026 Grenoble, France,}
\affiliation{$^{5}$Centro de Investigaciones Energ\'{e}ticas, Medioambientales Tecnol\'{o}gicas, E-28040 Madrid, Spain}
\begin{abstract}
The antiparticle energy spectra of Galactic cosmic rays (CRs) have several exciting features such as the 
unexpected positron excess at $E\sim$10-200\,GeV and the remarkably hard antiproton flux at $E\sim$\,60--450\,GeV 
recently measured by the \AMS{} experiment.
In this paper, we report calculations of antiparticle CR spectra arising from secondary production and their corresponding uncertainties.
Using the most recent data on CR protons, helium, carbon, and nuclear ratios $^{10}$Be/$^{9}$Be and B/C, 
we have performed a global Bayesian analysis, based on a Markov Chain Monte Carlo sampling algorithm,
under a scenario of spatial-dependent CR diffusion in the Galaxy which reproduces well the 
observed spectral hardening in the CR hadron fluxes.
While the high-energy positron excess requires the contribution of additional unknown sources,
we found that the antiproton data are consistent within the estimated uncertainties, with our predictions based on secondary production.
\end{abstract}
\maketitle

%%%%%%%%%%%%%%%%%%%%%%%%%%%%%%
\section{Introduction}\label{Sec::Introduction} %%%
%%%%%%%%%%%%%%%%%%%%%%%%%%%%%%

%In this paper we are mostly focused on CR antiparticle spectra. 
%According to the basic predictions of standard diffusion models, 
In standard models of CR propagation, antimatter particles are only produced by by collisions of high-energy 
nuclei with the interstellar gas, from which the \pbarp{} ratio or the positron fraction \epfrac{} are 
naively expected to \emph{decrease} with energy as fast as the boron-to-carbon (\BC) ratio does. 
To interpret the positron excess, it seems to be unavoidable to introduce extra source 
components such as dark matter particle annihilation
\cite{Cirelli:2008jk}, nearby supernova remnants \cite{Tomassetti:2015cva} 
or $e^{\pm}$ pair production mechanisms inside nearby pulsars 
\cite{Feng:2015uta}. 
The new \AMS{} data on the \pbarp{} ratio are also at tension with standard predictions based on secondary antiproton production.
However the situation is far from being understood % in the CR physics landscape,
%But the situation is not clear with CR diffusion which is far to be understood,
because conventional models of CR propagations suffers from large uncertainties and 
intrinsic limitations in describing the fine structures of new CR observations. 
For instance, recent data of CR proton and nuclei spectra revelead a high-energy departures 
from the standard universal power-law expectations. These features may challenge the
basic assumptions of standard diffusion models and give rise to exciting questions on
%gave rise to new speculations and exciting questions about 
how, where, and why CR propagation takes place \cite{Blasi:2013rva}. % cite Blasi review - 2013? JF added ref.
Among unjustified assumptions of traditional models, the homogenity of CR propagation has been
recently questioned. 
%JF: Johannesson2016 points out that the propagation parameters for p, p_bar and He is different from those of B-C-N-O. This is what they call "inhomogeneus".

%

%In this paper, we provide the \pbarp{} secondary production calculations under a two-halo model (THM) scenario
In this paper, we report the results of a complete scan of the parameter space for CR injection and propagation 
%based on a two-halo model (THM) scenario 
in order to provide a robust prediction for %the antiproton flux and % JF: Should we make a plot of flux > 45 GV ? Or we don't show pbar flux at all? 
the \pbarp{} ratio and corresponding uncertainties.
To describe the CR transport in the Galaxy, we set up a numerical implementation of a two-halo (THM) scenario of diffusive propagation
\cite{Tomassetti:2012ga, Tomassetti:2015mha, Guo:2015csa}, %add Jin Chao JF added ref.
where CRs are allowed to experience a different type of diffusion when they propagate closer to the Galactic plane. 
%We perform numerical calculations under a THM scenario which was proven to describe features of high-energy CR hadron spectra. 
%In THM, the propagation halo is ideally divided into two different regions along the z-axis, inner and outer, 
%where CRs experience different types of diffusion.
%will suffer from different propagation effect. 
%
To asses the uncertainties on the CR transport, we adopt a Markov Chain Monte Carlo (MCMC) sampling technique 
\cite{Lewis:2002ah},
using a large set of nuclear data (H, He, C, $^{10}$Be/$^{9}$Be, B/C),
%We use proton and other light-nuclei data (He, C, Be-10/Be-9, B/C) to determine the relevant parameters and their uncertainties. 
which allows us to inspect parameter correlations and their degeneracies.
We also review uncertainties associated to solar modulation and antiproton production cross sections.
We found that the new $\pbarp$ ratio measured by \AMS{} is fairly well described by a THM propagation model within the estimated errors,
while the excess of $\eplus$ require the presence of extra sources. We also give an example of one-zone diffusion models for comparison.

In Section\,\ref{Sec::Propagation_model} we give a description of the CR propagation model, 
the key parameters subjected to investigatoin and the method used for their determination.
In Section~\ref{Sec::Results} we report our results, we discuss the probability distribution 
inferred on the parameters and their degeneracy, and the overall description of CR data. 
We also present antiparticle spectra arising from secondary productions. 
The conclusions are drawn in Section~\ref{Sec::Conclusions}. 
%

%%%%%%%%%%%%%%%%%%%%%%%%%%%%%%%%%%%%%%%%%%%%%%%%%%%%%%%%%%%%%%%%%%%%%%%%%%%%%%%%%%%%%%%%%%%%%%%%
\section{Cosmic ray propagation model }\label{Sec::Propagation_model}  %%%
%%%%%%%%%%%%%%%%%%%%%%%%%%%%%%%%%%%%%%%%%%%%%%%%%%%%%%%%%%%%%%%%%%%%%%%%%%%%%%%%%%%%%%%%%%%%%%%%

%
%
%
The propagation of all CR species is described by a two-dimensional transport equation with 
boundary conditions at $r=r_{\rm max}$ and $z=\pm L$:
\begin{equation}\label{Eq::DiffusionTransport}
%  \partial_{t} {\psi} = Q + \vec{\nabla}\cdot (D\vec{\nabla}{\psi}) - {{\psi}}{\Gamma} + \partial_{E} (\dot{E} {\psi})  \,,
\frac{\partial {\psi}}{\partial t} = Q + \vec{\nabla}\cdot (D\vec{\nabla}{\psi}) - {{\psi}}{\Gamma} + \frac{\partial}{\partial E} (\dot{E} {\psi})  \,,
\end{equation}
where $\psi=\psi(E,r,z)$ is the particle number density as a function of energy and space coordinates,
$\Gamma= \beta c n \sigma$ is the destruction rate for collisions off gas nuclei, with density $n$,
at velocity $\beta c$ and cross section $\sigma$. The source term $Q$ is split into a primary term, $Q_{\rm pri}$, 
and a secondary production term $Q_{\rm sec}= \sum_{\rm j} \Gamma_{j}^{\rm sp} \psi_{\rm j}$, from spallation of 
heavier $j$--type nuclei with rate $\Gamma_{j}^{\rm sp}$. 
The term $\dot{E}=\--\frac{dE}{dt}$ describes ionization and Coulomb losses, as well as radiative cooling of CR leptons.

In this work, we adopt a spatial dependent scenario of CR diffusion in two halos.
In practice, this is made by ideally splitting the cylintrical propagation region into two $z$-symmetric halos
characterized by different diffusion coefficients.
We adopt a diffusion coefficient of the following form:
\begin{equation}\label{Eq::DiffusionCoefficient}
D_{xx}(\R,r,z) =  
\left\{
\begin{array}{r@{\; \;}l}
 D_0 \beta^{\eta}{\left(\frac{\R}{\R_0}\right)}^{\delta}  \quad\quad\quad&  (|z| < \xi L) \\
 \chi D_0 \beta^{\eta}{\left(\frac{\R}{\R_0}\right)}^{\delta F(r,z)} &  (\xi L < |z| < L) 
\end{array}
\right.
\end{equation}
where $F(r,z)$ is used to set a smooth transition between the two diffusion regions.
The parameter $D_0$ sets the normalization of the diffusion coefficient in the disk, 
at the reference rigidity $\R_{0}\equiv 0.25$\,GV, 
while $\chi D_0$ is that in the outer halo. The low-energy diffusion is shaped by the factor $\beta^{\eta}$, 
where $\eta$ is set to be $-0.4$ \cite{Gaggero:2013nfa}. The parameter $\delta$ represents the scale index of 
the power-law dependence of the diffusion coeffient in the inner halo (with $|z|< \xi L$) 
while $\delta+\Delta$ is that of the outer halo ($\xi L < |z|<L$), where $L$ is the half-height of the whole diffusion region.
%, while $\xi L$ is that of  the inner halo. 
%

In this work, we use the data to constraint the parameters
of CR transport in the two regions, namely 
$D_0$, $\chi$, $\delta$, $\Delta$, $L$ and $\xi$.
We also employ the parameter $\nu$ describing the proton injection index, and $\Delta\nu$ for the injection index 
difference between proton and other primary nuclei (\He, \C, \Oxy, ... \Fe).
%

%%%%%%%%%%%%%%%%%%%%%%%%%%%%%%%%%%%%%%%%%%%%%%%%%%%%%%%%%%%%%%%%%%%%
%\section{Scanned Results of Parameters \label{Sec::Results_parameter}} %%
\section{Results}\label{Sec::Results} %%
%%%%%%%%%%%%%%%%%%%%%%%%%%%%%%%%%%%%%%%%%%%%%%%%%%%%%%%%%%%%%%%%%%%%

%%%%%%%%%%%%%%%%%%%%%%%%%%%%%%%%%%%%%%%%%%%%%%%%%%%%%%%%%%%%%
\begin{figure}[!t]
\includegraphics[width=0.5\textwidth]{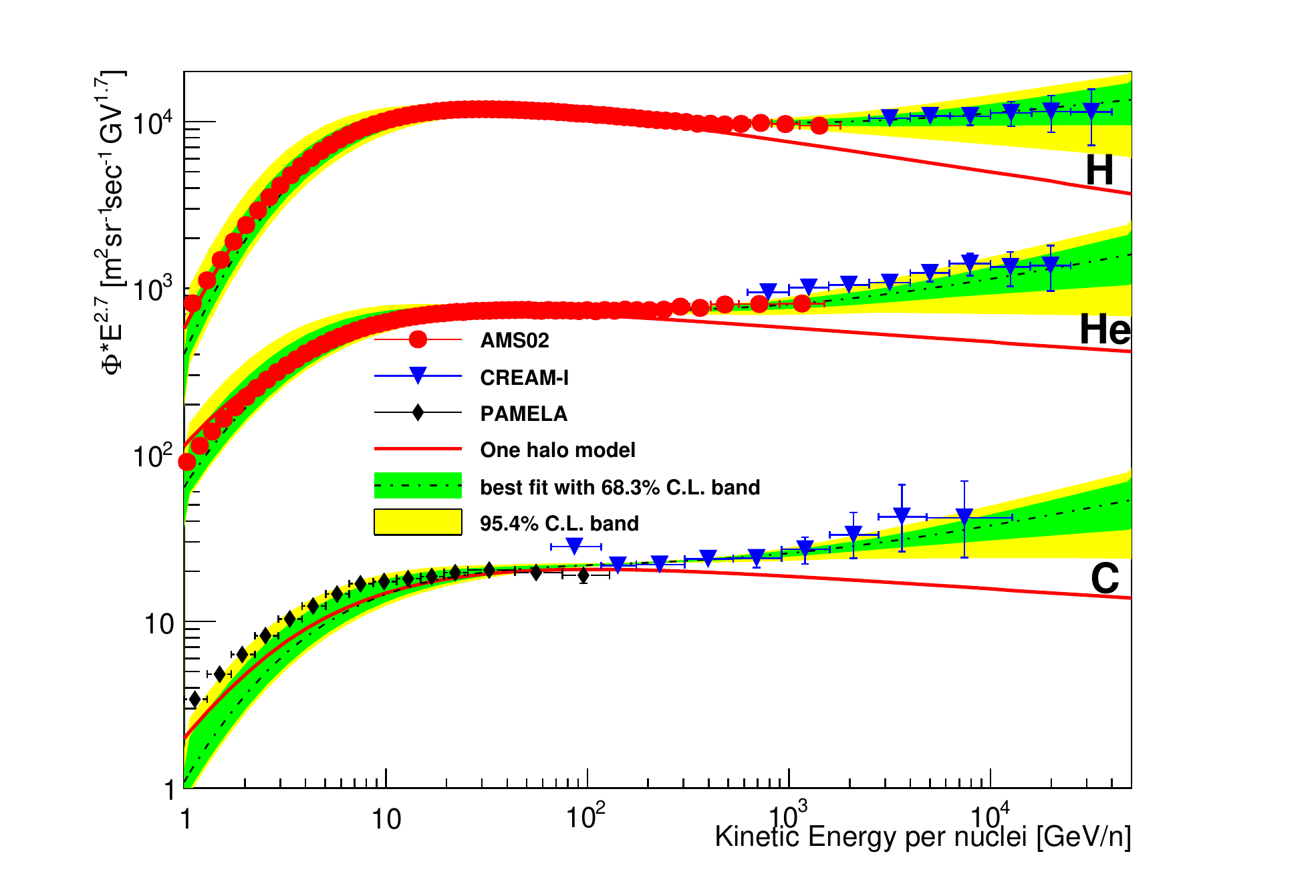} %_20
\includegraphics[width=0.5\textwidth]{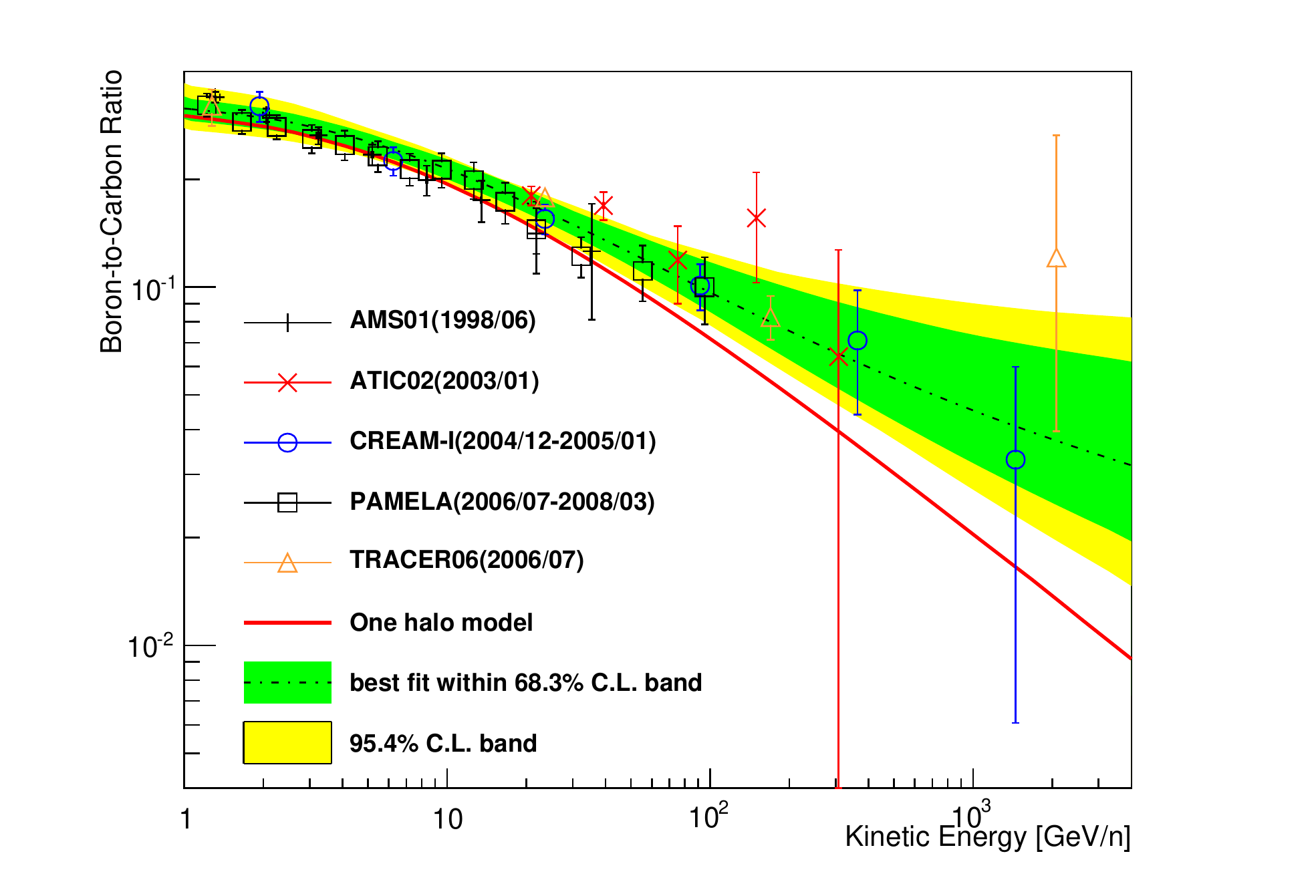}
\includegraphics[width=0.5\textwidth]{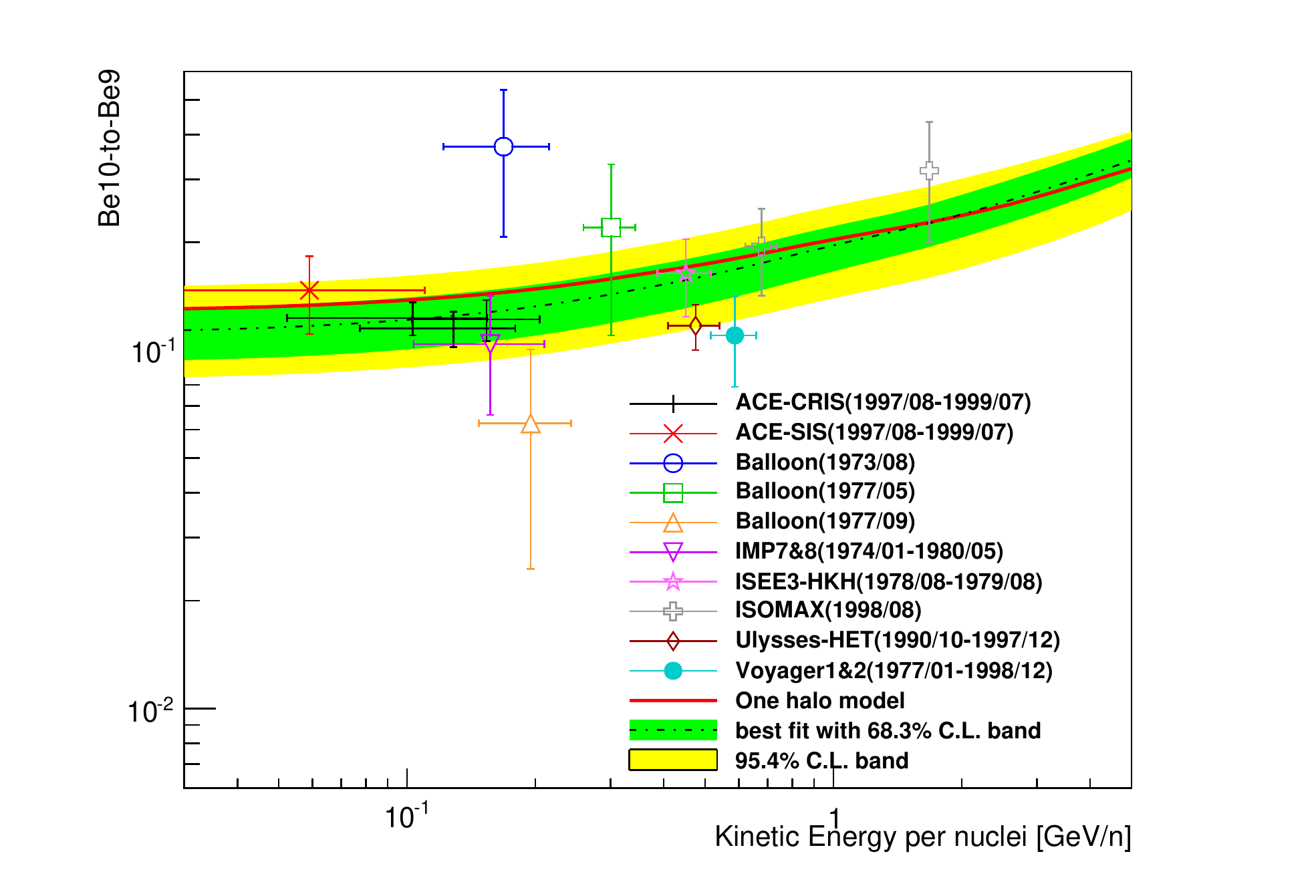}
\caption{
Top: Model prediction of proton, \He\ and \C\ fluxes compared with data. 
Middle: Prediction of \BC\ compared with data.
Bottom: Prediction of \BeBe\ compared with data.
} 
\label{Fig::ccProtonSpectrum}%
\end{figure}
%%%%%%%%%%%%%%%%%%%%%%%%%%%%%%%%%%%%%%%%%%%%%%%%%%%%%%%%%%%%%
%%%%%%%%%%%%%%%%%%%%%%%%%%%%%%%%%%%%%%%%%%%%%%%%%%%%%%%%%%%%%
\begin{figure}[!t]
\includegraphics[width=0.44\textwidth]{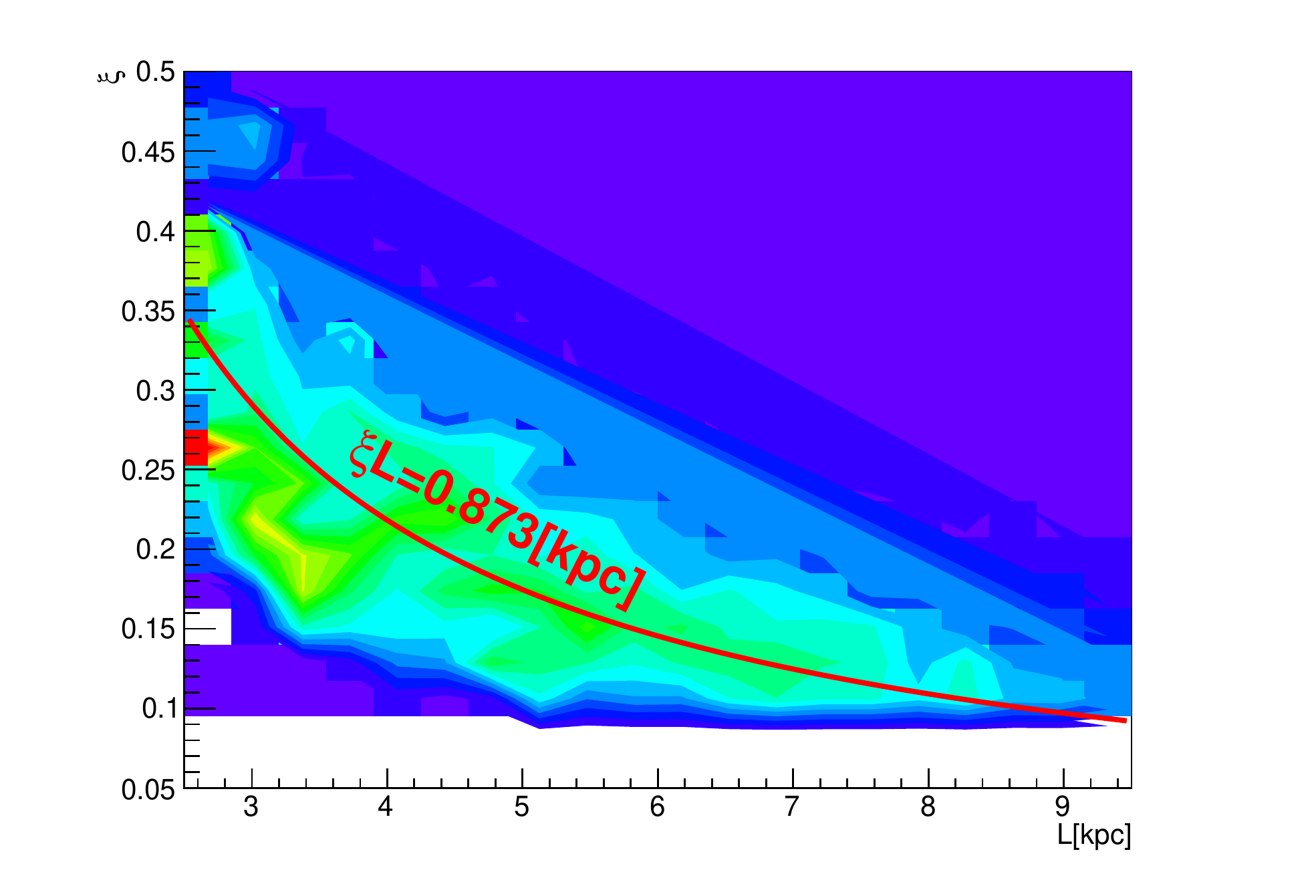}\caption{
	Contour in the $\xi$ and $L$ plane. The red solide line at $\xi \times L=0.873$\,[kpc] is to guid the eye.}
\label{Fig::xi_L}%
\end{figure}
%%%%%%%%%%%%%%%%%%%%%%%%%%%%%%%%%%%%%%%%%%%%%%%%%%%%%%%%%%%%%
%%%%%%%%%%%%%%%%%%%%%%%%%%%%%%%%%%%%%%%%%%%%%%%%%%%%%%%
\begin{figure}[!t]
\includegraphics[width=0.44\textwidth]{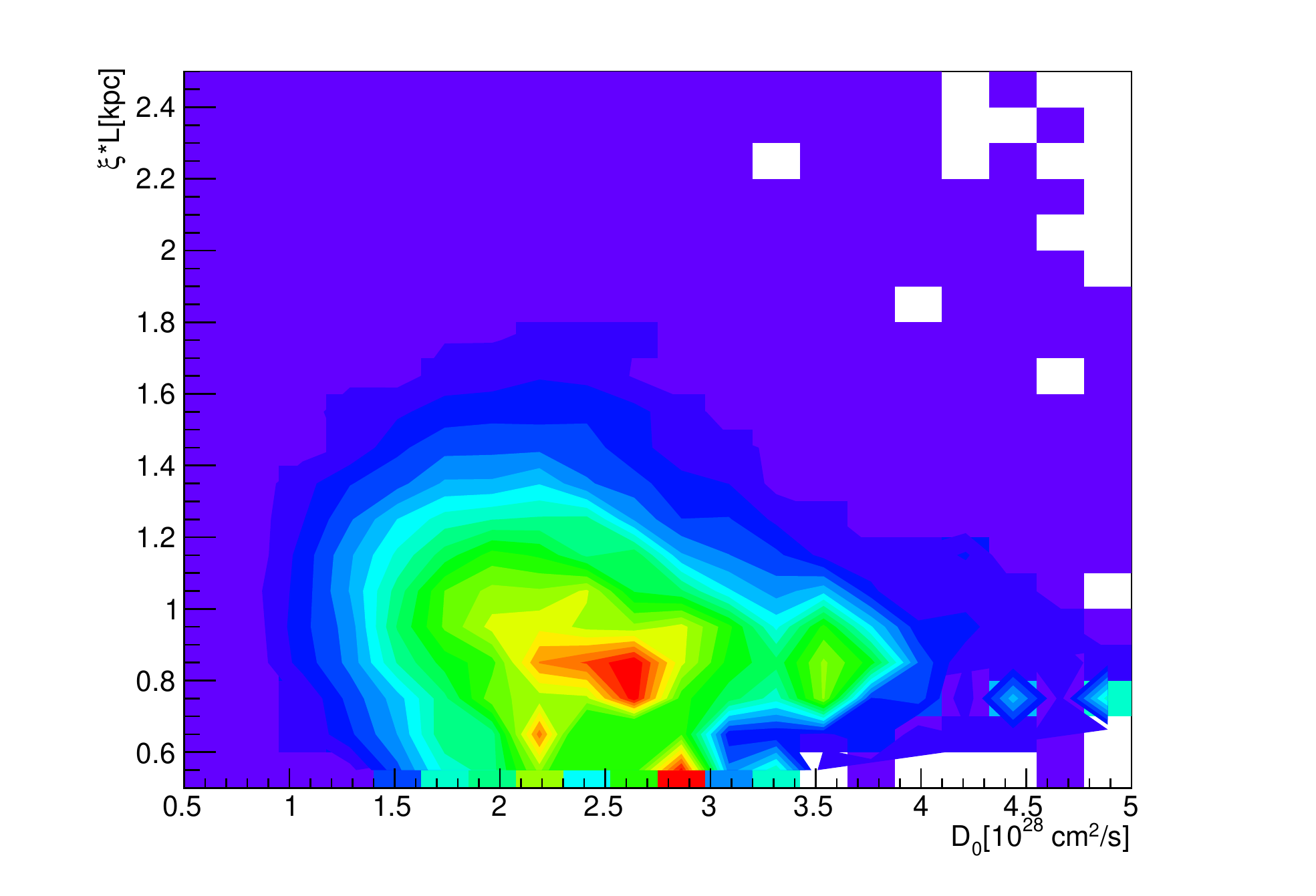}
\includegraphics[width=0.44\textwidth]{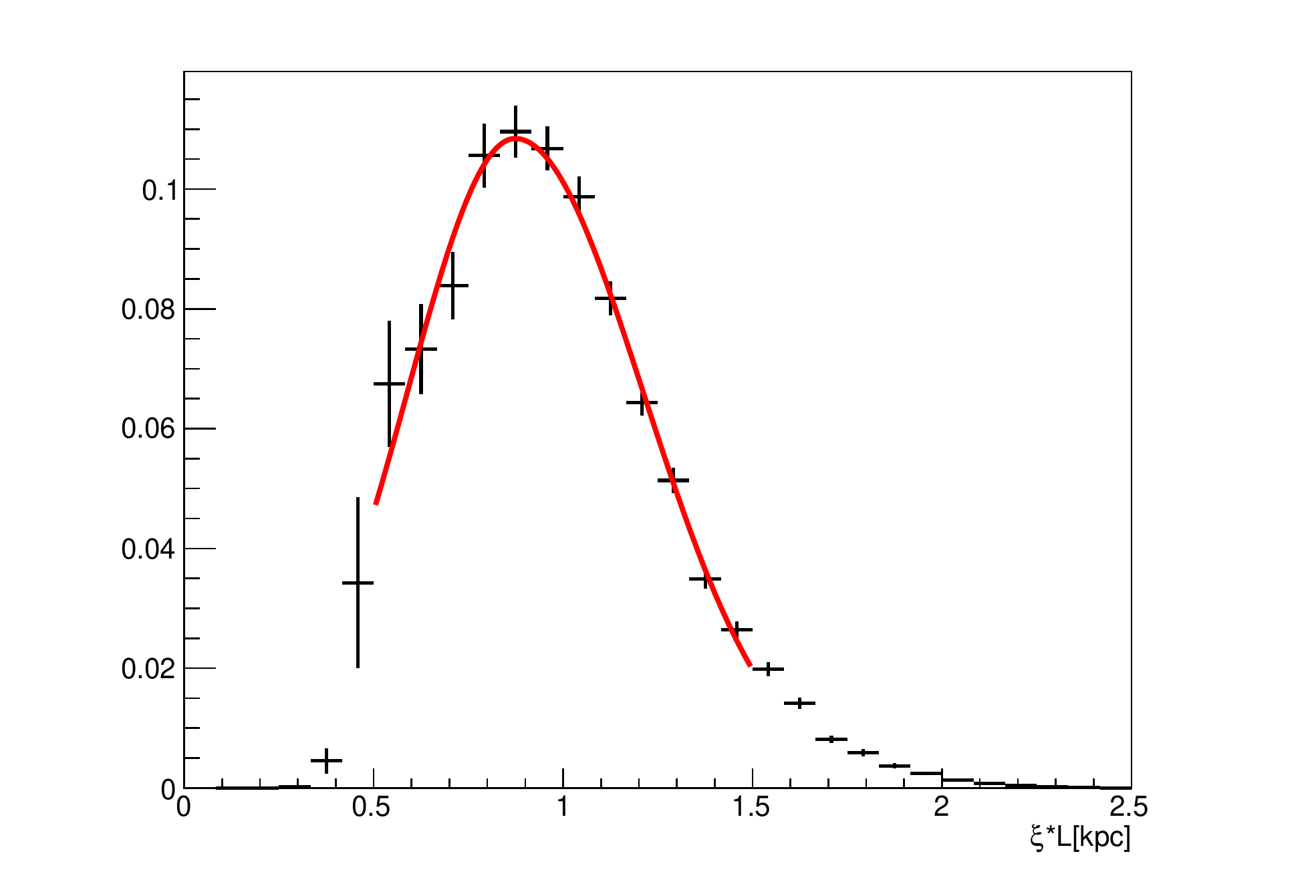}
\caption{
Top: Contour in $\xi$$L$ and $D_0$ plane. Buttom: distribution of $\xi$$L$.
}
\label{Fig::xiL}
\end{figure}
%%%%%%%%%%%%%%%%%%%%%%%%%%%%%%%%%%%%%%%%%%%%%%%%%%%%%%%
%%%%%%%%%%%%%%%%%%%%%%%%%%%%%%%%%%%%%%%%%%%%%%%%%%%%%%%%%%%%%
\begin{figure}[!t]
\includegraphics[width=0.44\textwidth]{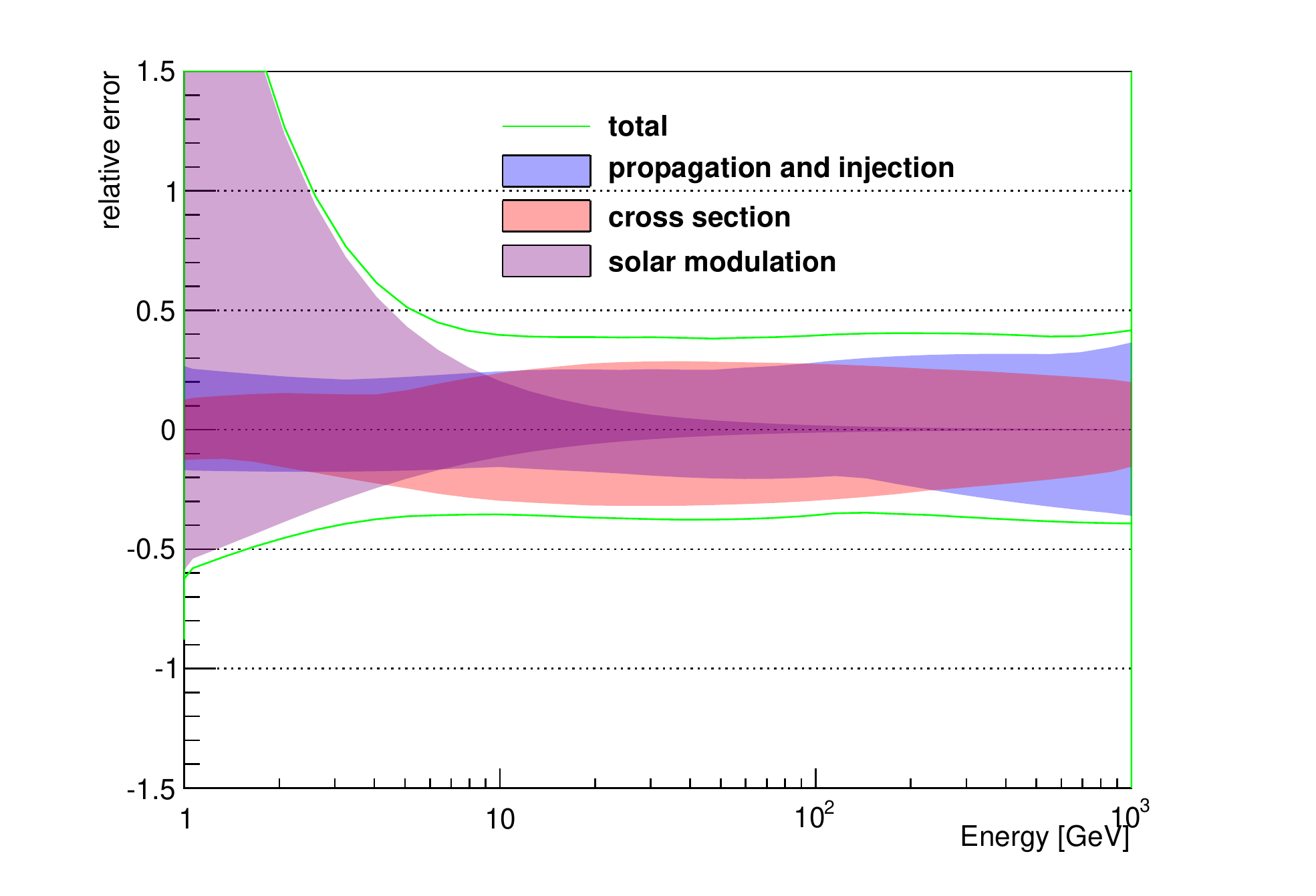}
\includegraphics[width=0.44\textwidth]{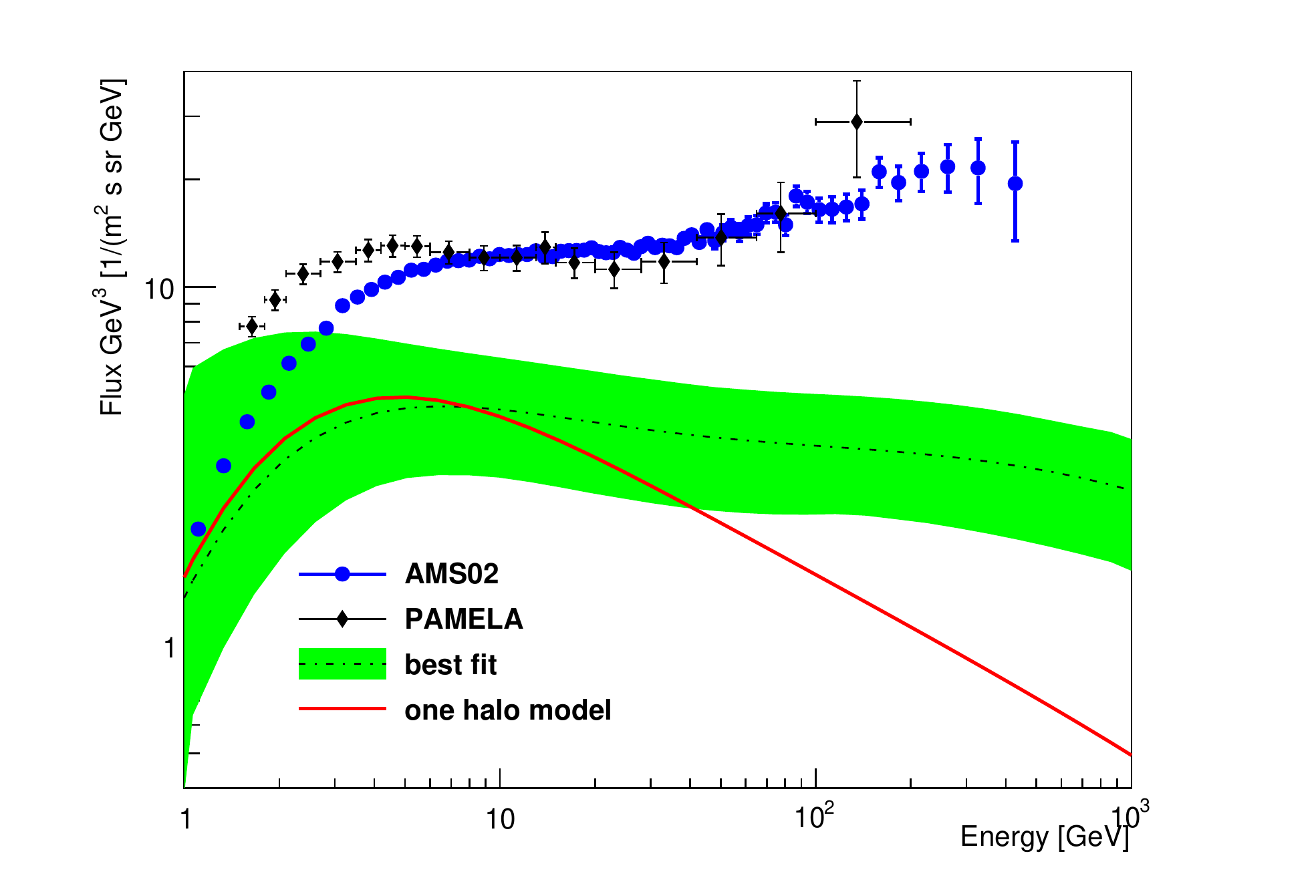}\caption{
Top: $e+$ flux prediction error break down.
Buttom:	Model prediction of positron flux compared with the experimental data by \AMS\ \cite{Aguilar:2014mma} as well as PAMELA \cite{Adriani:2013uda}.}
\label{Fig::positrons}%
\end{figure}
%%%%%%%%%%%%%%%%%%%%%%%%%%%%%%%%%%%%%%%%%%%%%
%%%%%%%%%%%%%%%%%%%%%%%%%%%%%%%%%%%%%%%%%%%%%%%%%%%%%%%
\begin{figure}[!t]
\includegraphics[width=0.44\textwidth]{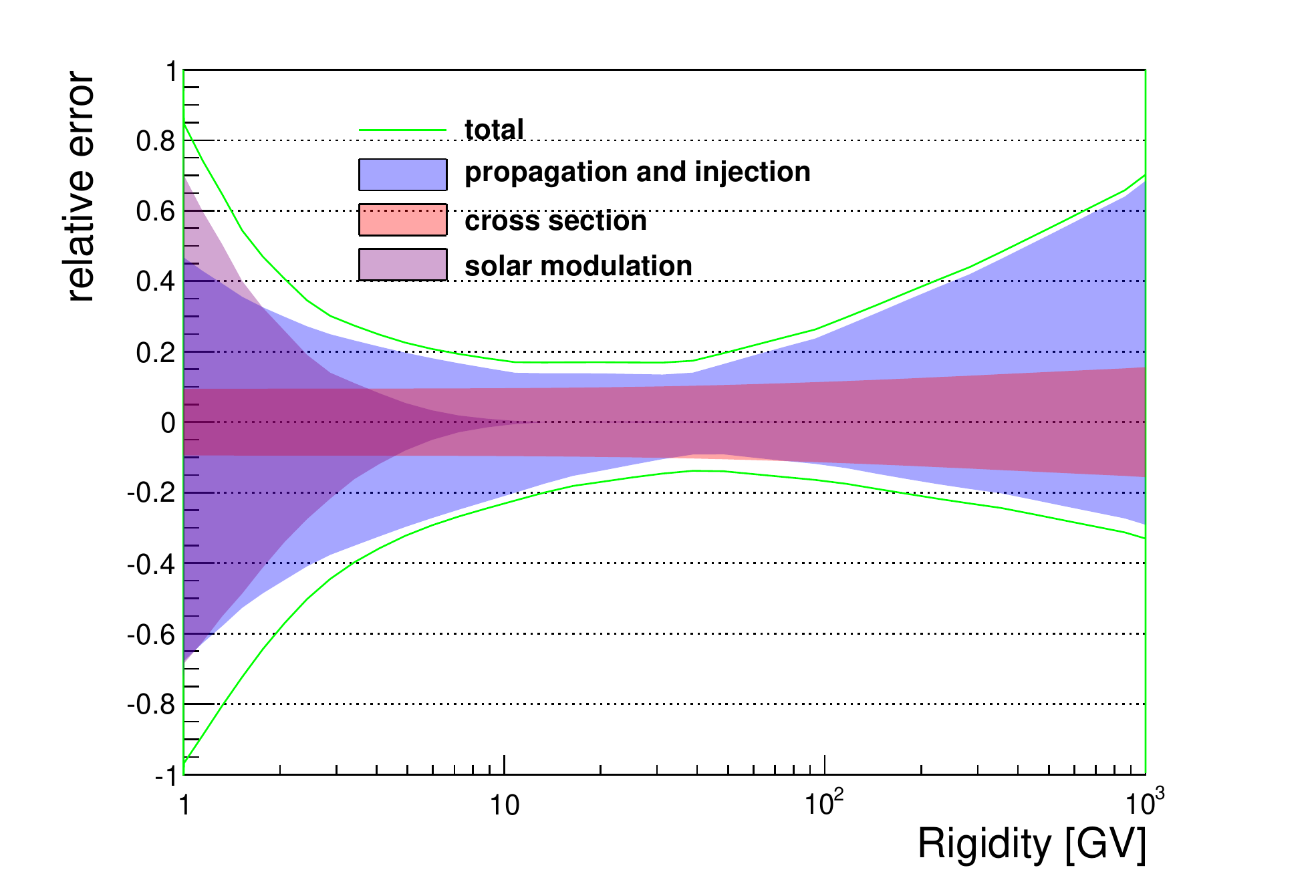}
\includegraphics[width=0.44\textwidth]{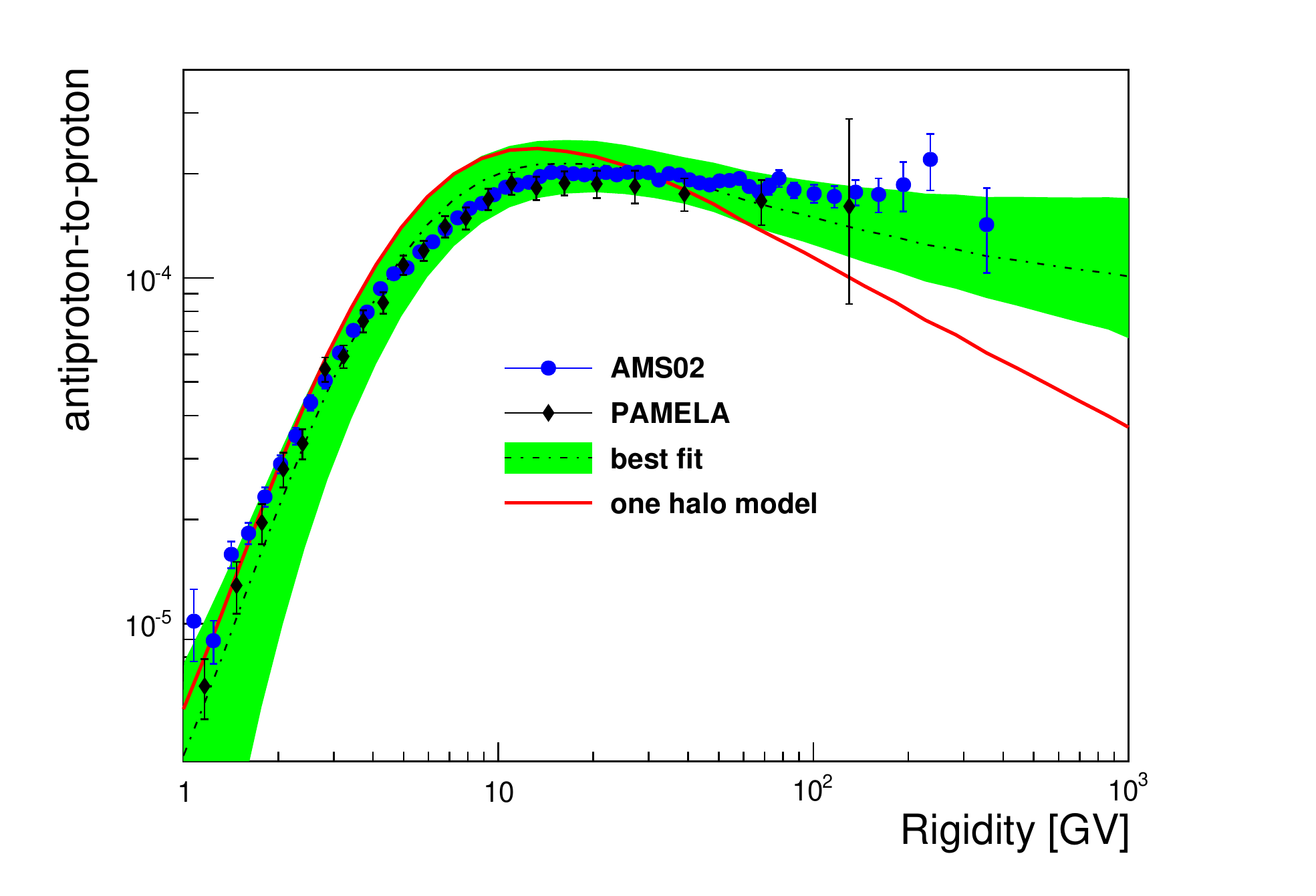}
\caption{
Top: $\bar{p}$ prediction error break down. Buttom: 
  The \pbarp{} ratio as function of rigidity $R$.
  The model calculations
  are shown in comparison with the \AMS{} \cite{PhysRevLett.117.091103} and PAMELA \cite{Adriani:2012paa} data. 
}
\label{Fig::error}
\end{figure}

%%%%%%%%%%%%%%%%%%%%%%%%%%%%%%%%%%%%%%%%%%%%%
The best fit parameters can reproduce the primary spectra very well.
Figure\,\ref{Fig::ccProtonSpectrum} shows the spectra of  of proton, \He{} and \C,
where the THM fit is plotted as dashed lines and the shaded areas 
represent the 1-$\sigma$ (green) and 2-$\sigma$ (yellow) uncertainty bands. 
Our results are nicely consistent with the trends obtained in former analytical derivations \cite{Tomassetti:2012ga}:
following the trend of the data, the spectra get progressively harder at energies above $\sim$\,200\,GeV/n.
In the figure, the thin solid lines represents the prediction from the standard one-halo (OHM) scenario, shown for reference,
which is fitted to data at energie $E\lesssim$\,200 GeV/n. 

The top plot of Fig.\,\ref{Fig::ccProtonSpectrum} shows us that THM predicts the primary particle spectra with the same quantity of deviation from a 
single pow-law spectra. This is expected because the primary particles propagate in the disk and halo results in $\phi_{p}\sim Q/D$, with $\frac{1}{D}=
\frac{L}{D_{0}R^{\delta}}(\xi+\frac{1}{\chi R^{\Delta}})$. The middle plot of Fig.\,\ref{Fig::ccProtonSpectrum}, however, shows us that the secondary 
particle spectra has a larger high energy break than primary ones: a larger break of \B\ than that of \C\ results in the hardening behavior in \BC. 
This has also been pointed out in \cite{Tomassetti:2015mha}. 
One should also notice that nuclear fragmentation cross section will introduce uncertainties \cite{Tomassetti:2015XS, Tomassetti:2012zb}.

The correlation between  $\xi$ and $L$ is also shown in the contour plot of Fig.\,\ref{Fig::xi_L}.
Even though we have included \BeBe{} data in our study, the size of halo $L$ does not converge. %Hmm CITARE PUTZE2010 ?

This is illustrated in Fig.~\ref{Fig::xi_L}, showing the convergence of the inner halo height $\xi$$L$ to the value of $\sim$\,1\,kpc. 
In Fig.~\ref{Fig::xiL} we directly plot the contour in the ($\xi$$L$ - $D_0$) plane: it can be see that no significan correlation
is present between $\xi$$L$ and $D_0$. Also, we have performed a fit on the $\xi$$L$ distribution in the interval
$\xi$$L\in \left[0.5, 1.5\right]$ (kpc) with an asymmetric gaussian function, as shown in Fig.\,\ref{Fig::xiL}. 
The fit discribes the distribution quite well with $\chi^{2}/d.f. = 8.15/8$, with a mean value of $(0.873^{+0.340}_{-0.286})$\,kpc.

Positron flux predicted by THM is shown in Fig.~\ref{Fig::positrons}. The positron flux predicted by THM is harder than that by OHM. One can expect 
this because positron spend a larger fraction of time in the disk in THM where the $\delta$ is smaller than that in OHM. 
We give a break down of the errors from the model. The cross section error is estimated as in \cite{Delahaye:2008ua}.

The antiprotons emitted in \p-\p(ISM), \p-\He(ISM), \He-\p(ISM) and \He-\He(ISM) collisions are characterized
by broad energy distributions and large inelasticity factors. Antineutrons produced from the above collisions also decay to become antiprotons.
The transport properties of antiprotons in the Galaxy is similar to that of protons. However, non-annihilating reactions of
secondary $\bar{p}$ with the gas may also produce a further energy shift toward lower energies.
Comparing the MC generator models with the most recent experimental data \cite{Feng:2016loc}, we validate EPOS LHC \cite{Pierog:2013ria}  
as the best one and use it in this work.
Apart from the propagation and cross section errors discussed above, we also need to consider solar modulation error especially 
for low energies. 
Comparing to the antiproton flux, $\bar{p}/p$ is a better observable to eliminate most of the solar modulation and the proton normalization uncertainties. 
The top graph of Fig. 
~\ref{Fig::error} shows the break down of the $\bar{p}/p$ prediction errors. 

The \pbarp{} ratio is shown for the two considered models in the bottom graph of Fig.\,\ref{Fig::error}. In OHM, it decreases smoothly above 
$\sim$\, 10\,GeV. In our THM, the ratio has a flattening tendency at $E\sim$\,10\,GeV to $\sim$\, 100\,GeV. 

%%%%%%%%%%%%%%%%%%%%%%%%%%%%%
\section{Conclusions}     %%%
\label{Sec::Conclusions}  %%%
%%%%%%%%%%%%%%%%%%%%%%%%%%%%%

Here we summarize everything. 
We estimate the inner halo size, namely disk height, of our galaxy to be around 1 kpc. The outer halo size should be further studied.
The Bayesian analysis gives us the uncertainty band from propagation and nuclei injection. We also study the antiproton production 
cross sections carefully and validate EPOS LHC model. We show the error break down of the antiparticle predictions. The uncertainties 
from \pbarp{} measurement by \AMS{} are already smaller than those from model predictions. It is shown that there is no room for extra 
source to produce antiprotons. Positron prediction shows that there is a hint of the existence of extra sources which produce or 
accelerate positrons but not secondary nuclei.  
%
%
%%%%%%%%%%%%%%%%%%%%%%%%%%%%%%%
\section*{Acknowledgments}  %%%
%%%%%%%%%%%%%%%%%%%%%%%%%%%%%%%

   JF acknowledges the Taiwanese Ministry of Science and
   Technology (MOST) under Grant No. 104-2112-M-001-024 and Grant No. 105-2112-M-001-003.
   AO acknowledges CIEMAT, CDTI and SEIDI MINECO under Grants ESP2015-71662-C2-(1-P) and MDM-2015-0509.
   NT acknowledges support from MAtISSE - \emph{Multichannel Investigation of Solar Modulation Effects in Galactic Cosmic Rays}.
This project has received funding from the European Union's Horizon 2020 research and innovation programme under the Marie Sklodowska-Curie grant agreement No 707543.

%%%%%%%%%%%%%%%%%%%%%%%%%%%%%%%%%%%%%%%%%%%%%%%%%%%%%%%%%%%%%%%%%%%%%%%%%%%%%

%%%%%%%%%%%%%%%%%%%%%%%%%%%%%%%%%%%%%%%%%%%%%%%%%%%%%%%%%%%%
%%%%%%%%%%%%%%%%%%%%%%%%%%%%%%%%%%%%%%%%%%%%%%%%%%%%%%%
\bibliography{thm_mcmc_nt_jf}

\newcommand{\aap}{A\&A} \newcommand{\mnras}{Monthly Notices of the Royal
  Astronomical Society} \newcommand{\sovast}{Soviet Astronomy}
\begin{thebibliography}{18}
\expandafter\ifx\csname natexlab\endcsname\relax\def\natexlab#1{#1}\fi
\expandafter\ifx\csname bibnamefont\endcsname\relax
  \def\bibnamefont#1{#1}\fi
\expandafter\ifx\csname bibfnamefont\endcsname\relax
  \def\bibfnamefont#1{#1}\fi
\expandafter\ifx\csname citenamefont\endcsname\relax
  \def\citenamefont#1{#1}\fi
\expandafter\ifx\csname url\endcsname\relax
  \def\url#1{\texttt{#1}}\fi
\expandafter\ifx\csname urlprefix\endcsname\relax\def\urlprefix{URL }\fi
\providecommand{\bibinfo}[2]{#2}
\providecommand{\eprint}[2][]{\url{#2}}

\bibitem[{\citenamefont{{Cirelli} and {Strumia}}(2008)}]{Cirelli:2008jk}
\bibinfo{author}{\bibfnamefont{M.}~\bibnamefont{{Cirelli}}} \bibnamefont{and}
  \bibinfo{author}{\bibfnamefont{A.}~\bibnamefont{{Strumia}}},
  \bibinfo{journal}{ArXiv e-prints}  (\bibinfo{year}{2008}),
  \eprint{0808.3867}.

\bibitem[{\citenamefont{Tomassetti and Donato}(2015)}]{Tomassetti:2015cva}
\bibinfo{author}{\bibfnamefont{N.}~\bibnamefont{Tomassetti}} \bibnamefont{and}
  \bibinfo{author}{\bibfnamefont{F.}~\bibnamefont{Donato}},
  \bibinfo{journal}{Astrophys.J.} \textbf{\bibinfo{volume}{803}},
  \bibinfo{pages}{L15} (\bibinfo{year}{2015}), \eprint{1502.06150}.

\bibitem[{\citenamefont{Feng and Zhang}(2016)}]{Feng:2015uta}
\bibinfo{author}{\bibfnamefont{J.}~\bibnamefont{Feng}} \bibnamefont{and}
  \bibinfo{author}{\bibfnamefont{H.-H.} \bibnamefont{Zhang}},
  \bibinfo{journal}{Eur. Phys. J.} \textbf{\bibinfo{volume}{C76}},
  \bibinfo{pages}{229} (\bibinfo{year}{2016}), \eprint{1504.03312}.

\bibitem[{\citenamefont{Blasi}(2013)}]{Blasi:2013rva}
\bibinfo{author}{\bibfnamefont{P.}~\bibnamefont{Blasi}},
  \bibinfo{journal}{Astron. Astrophys. Rev.} \textbf{\bibinfo{volume}{21}},
  \bibinfo{pages}{70} (\bibinfo{year}{2013}), \eprint{1311.7346}.

\bibitem[{\citenamefont{Tomassetti}(2012{\natexlab{a}})}]{Tomassetti:2012ga}
\bibinfo{author}{\bibfnamefont{N.}~\bibnamefont{Tomassetti}},
  \bibinfo{journal}{Astrophys. J.} \textbf{\bibinfo{volume}{752}},
  \bibinfo{pages}{L13} (\bibinfo{year}{2012}{\natexlab{a}}),
  \eprint{1204.4492}.

\bibitem[{\citenamefont{Tomassetti}(2015{\natexlab{a}})}]{Tomassetti:2015mha}
\bibinfo{author}{\bibfnamefont{N.}~\bibnamefont{Tomassetti}},
  \bibinfo{journal}{Phys. Rev. D} \textbf{\bibinfo{volume}{92}},
  \bibinfo{pages}{081301} (\bibinfo{year}{2015}{\natexlab{a}}),
  \eprint{1509.05775}.

\bibitem[{\citenamefont{Guo et~al.}(2015)\citenamefont{Guo, Tian, and
  Jin}}]{Guo:2015csa}
\bibinfo{author}{\bibfnamefont{Y.-Q.} \bibnamefont{Guo}},
  \bibinfo{author}{\bibfnamefont{Z.}~\bibnamefont{Tian}}, \bibnamefont{and}
  \bibinfo{author}{\bibfnamefont{C.}~\bibnamefont{Jin}} (\bibinfo{year}{2015}),
  \eprint{1509.08227}.

\bibitem[{\citenamefont{Lewis and Bridle}(2002)}]{Lewis:2002ah}
\bibinfo{author}{\bibfnamefont{A.}~\bibnamefont{Lewis}} \bibnamefont{and}
  \bibinfo{author}{\bibfnamefont{S.}~\bibnamefont{Bridle}},
  \bibinfo{journal}{Phys. Rev. D} \textbf{\bibinfo{volume}{66}},
  \bibinfo{pages}{103511} (\bibinfo{year}{2002}), \eprint{astro-ph/0205436}.

\bibitem[{\citenamefont{Gaggero et~al.}(2014)\citenamefont{Gaggero, Maccione,
  Grasso, Di~Bernardo, and Evoli}}]{Gaggero:2013nfa}
\bibinfo{author}{\bibfnamefont{D.}~\bibnamefont{Gaggero}},
  \bibinfo{author}{\bibfnamefont{L.}~\bibnamefont{Maccione}},
  \bibinfo{author}{\bibfnamefont{D.}~\bibnamefont{Grasso}},
  \bibinfo{author}{\bibfnamefont{G.}~\bibnamefont{Di~Bernardo}},
  \bibnamefont{and} \bibinfo{author}{\bibfnamefont{C.}~\bibnamefont{Evoli}},
  \bibinfo{journal}{Phys. Rev. D} \textbf{\bibinfo{volume}{89}},
  \bibinfo{pages}{083007} (\bibinfo{year}{2014}), \eprint{1311.5575}.

\bibitem[{\citenamefont{Aguilar et~al.}(2014)}]{Aguilar:2014mma}
\bibinfo{author}{\bibfnamefont{M.}~\bibnamefont{Aguilar}} \bibnamefont{et~al.}
  (\bibinfo{collaboration}{AMS}), \bibinfo{journal}{Phys.Rev.Lett.}
  \textbf{\bibinfo{volume}{113}}, \bibinfo{pages}{121102}
  (\bibinfo{year}{2014}).

\bibitem[{\citenamefont{Adriani et~al.}(2013{\natexlab{a}})}]{Adriani:2013uda}
\bibinfo{author}{\bibfnamefont{O.}~\bibnamefont{Adriani}} \bibnamefont{et~al.}
  (\bibinfo{collaboration}{PAMELA}), \bibinfo{journal}{Phys. Rev. Lett.}
  \textbf{\bibinfo{volume}{111}}, \bibinfo{pages}{081102}
  (\bibinfo{year}{2013}{\natexlab{a}}), \eprint{1308.0133}.

\bibitem[{\citenamefont{Aguilar et~al.}(2016)}]{PhysRevLett.117.091103}
\bibinfo{author}{\bibfnamefont{M.}~\bibnamefont{Aguilar}} \bibnamefont{et~al.}
  (\bibinfo{collaboration}{AMS Collaboration}), \bibinfo{journal}{Phys. Rev.
  Lett.} \textbf{\bibinfo{volume}{117}}, \bibinfo{pages}{091103}
  (\bibinfo{year}{2016}),
  \urlprefix\url{http://link.aps.org/doi/10.1103/PhysRevLett.117.091103}.

\bibitem[{\citenamefont{Adriani et~al.}(2013{\natexlab{b}})}]{Adriani:2012paa}
\bibinfo{author}{\bibfnamefont{O.}~\bibnamefont{Adriani}} \bibnamefont{et~al.},
  \bibinfo{journal}{JETP Lett.} \textbf{\bibinfo{volume}{96}},
  \bibinfo{pages}{621} (\bibinfo{year}{2013}{\natexlab{b}}),
  \bibinfo{note}{[Pisma Zh. Eksp. Teor. Fiz.96,693(2012)]}.

\bibitem[{\citenamefont{Tomassetti}(2015{\natexlab{b}})}]{Tomassetti:2015XS}
\bibinfo{author}{\bibfnamefont{N.}~\bibnamefont{Tomassetti}},
  \bibinfo{journal}{Phys. Rev. C} \textbf{\bibinfo{volume}{92}},
  \bibinfo{pages}{045808} (\bibinfo{year}{2015}{\natexlab{b}}),
  \urlprefix\url{http://link.aps.org/doi/10.1103/PhysRevC.92.045808}.

\bibitem[{\citenamefont{Tomassetti}(2012{\natexlab{b}})}]{Tomassetti:2012zb}
\bibinfo{author}{\bibfnamefont{N.}~\bibnamefont{Tomassetti}},
  \bibinfo{journal}{Astrophys. Space Sci.} \textbf{\bibinfo{volume}{342}},
  \bibinfo{pages}{131} (\bibinfo{year}{2012}{\natexlab{b}}),
  \eprint{1210.7355}.

\bibitem[{\citenamefont{Delahaye et~al.}(2009)\citenamefont{Delahaye, Donato,
  Fornengo, Lavalle, Lineros, Salati, and Taillet}}]{Delahaye:2008ua}
\bibinfo{author}{\bibfnamefont{T.}~\bibnamefont{Delahaye}},
  \bibinfo{author}{\bibfnamefont{F.}~\bibnamefont{Donato}},
  \bibinfo{author}{\bibfnamefont{N.}~\bibnamefont{Fornengo}},
  \bibinfo{author}{\bibfnamefont{J.}~\bibnamefont{Lavalle}},
  \bibinfo{author}{\bibfnamefont{R.}~\bibnamefont{Lineros}},
  \bibinfo{author}{\bibfnamefont{P.}~\bibnamefont{Salati}}, \bibnamefont{and}
  \bibinfo{author}{\bibfnamefont{R.}~\bibnamefont{Taillet}},
  \bibinfo{journal}{Astron. Astrophys.} \textbf{\bibinfo{volume}{501}},
  \bibinfo{pages}{821} (\bibinfo{year}{2009}), \eprint{0809.5268}.

\bibitem[{\citenamefont{Feng et~al.}(2016)\citenamefont{Feng, Tomassetti, and
  Oliva}}]{Feng:2016loc}
\bibinfo{author}{\bibfnamefont{J.}~\bibnamefont{Feng}},
  \bibinfo{author}{\bibfnamefont{N.}~\bibnamefont{Tomassetti}},
  \bibnamefont{and} \bibinfo{author}{\bibfnamefont{A.}~\bibnamefont{Oliva}},
  \bibinfo{journal}{Phys. Rev.} \textbf{\bibinfo{volume}{D94}},
  \bibinfo{pages}{123007} (\bibinfo{year}{2016}), \eprint{1610.06182}.

\bibitem[{\citenamefont{Pierog et~al.}(2015)\citenamefont{Pierog, Karpenko,
  Katzy, Yatsenko, and Werner}}]{Pierog:2013ria}
\bibinfo{author}{\bibfnamefont{T.}~\bibnamefont{Pierog}},
  \bibinfo{author}{\bibfnamefont{I.}~\bibnamefont{Karpenko}},
  \bibinfo{author}{\bibfnamefont{J.~M.} \bibnamefont{Katzy}},
  \bibinfo{author}{\bibfnamefont{E.}~\bibnamefont{Yatsenko}}, \bibnamefont{and}
  \bibinfo{author}{\bibfnamefont{K.}~\bibnamefont{Werner}},
  \bibinfo{journal}{Phys. Rev. C} \textbf{\bibinfo{volume}{92}},
  \bibinfo{pages}{034906} (\bibinfo{year}{2015}), \eprint{1306.0121}.

\end{thebibliography}
%%%%%%%%%%%%%%%%%%%%%%%%%%%%%%%%%%%%%%%%%%%%%%%%%%%%%%%
%%%%%%%%%%%%%%%%%%%%%%%%%%%%%%%%%%%%%%%%%%%%%%%%%%%%%%%
\end{document}